\documentclass[preprint,superscriptaddress,showpacs,floatfix,onecolumn]{revtex4-1}
\usepackage{graphicx,amsmath,amssymb,dsfont,bm,float,xcolor,ipa}
\usepackage{subcaption}

\usepackage[left=2.8cm,top=2.8cm,right=2.7cm,bottom=2.9cm,nohead]{geometry}

\definecolor{dgreen}{RGB}{0,127,10}
\usepackage[colorlinks=true,
            linkcolor=red,
            citecolor=dgreen,
            urlcolor=blue]{hyperref}

\newcommand{\tr}{\text{\,tr\,}}

\newcommand{\diag}{\text{diag\,}}

\newcommand{\1}{\mathds{1}_n}

\begin{document}
\widetext

\title{\Large Eigenvalue statistics for the sum of two complex Wishart matrices}

\author{Santosh Kumar} \email{skumar.physics@gmail.com} 
\affiliation{Department of Physics, Shiv Nadar University, Gautam Budh Nagar, Uttar Pradesh - 201314, India}

\begin{abstract}
The sum of independent Wishart matrices, taken from distributions with unequal covariance matrices, plays a crucial role in multivariate statistics, and has applications in the fields of quantitative finance and telecommunication. However, analytical results concerning the corresponding eigenvalue statistics have remained unavailable, even for the sum of two Wishart matrices. This can be attributed to the complicated and rotationally-noninvariant nature of the matrix distribution that makes extracting the information about eigenvalues a nontrivial task. Using a generalization of the Harish-Chandra-Itzykson-Zuber integral, we find exact solution to this problem for the case when one of the covariance matrices is proportional to the identity matrix, while the other is arbitrary. We find exact and compact expressions for the joint probability density and marginal density of eigenvalues. The analytical results are compared with numerical simulations and we find perfect agreement.
\end{abstract}

\pacs{02.50.Sk, 05.45.Tp, 89.90.+n, 05.45.Mt}
\maketitle

%\thispagestyle{empty}

% SECTION 1
\section{Introduction}
Wishart random matrices are named after John Wishart who worked out their distribution in 1928~\cite{Wishart1928}. Wishart distribution generalizes the $\chi^2$-distribution to the case of multiple variables. Since their inception, Wishart matrices have played a prominent role in the area of multivariate statistics~\cite{James1964,Muirhead2009,Anderson2003,AHSE1995,GN1999}. In recent years there has been a renewed and growing interest in their study because of their applicability in analyzing a variety of unrelated complex problems. For instance, on the one hand Wishart matrices have been implemented to analyze financial data~\cite{LCBM1999,PGRAGS2002,AFV2010,SCSG2014}. On the other hand they have been used to identify vulnerable regions in the human immunodeficiency virus (HIV), which could lead to effective AIDS vaccines or drugs~\cite{Detal2011}. Further examples, where Wishart matrices appear include telecommunication networks~\cite{Telatar1999,TV2004,SMM2006,ZCW2009,KP2010a,KP2011a}, quantum chromodynamics~\cite{SV1993,Verbaarschot1994,GW1997,VW2000,Akemann2011}, quantum entanglement problem~\cite{ZS2001,BL2002,NM2010,KP2011b,VZ2012}, mesoscopic systems~\cite{FH1994,SN1994}, gene expression data analysis~\cite{HMMCBF2000,ABB2000}, etc.

Random matrix ensembles involving various combinations of Wishart matrices are also relevant to several problems. The Jacobi or MANOVA (Multivariate ANalysis Of VAriance) ensemble is an example which is useful in the quantum conductance problem, and optical fiber communication studies~\cite{KP2011a,DE2002,Forrester2006,KP2010b,DFS2013}. Very recently several results involving the product of Wishart matrices have appeared in the literature~\cite{AIK2013,AKW2013}. These ensembles find applications in telecommunication of multi-layered scattering multiple-input and multiple-output channels. 

Another important ensemble which plays a crucial role in the multivariate statistics comprises sum of Wishart matrices~\cite{Anderson2003,AHSE1995,GN1999,Khatri1966,TG1983,NM1986,Sheppard2008}. They arise in matrix quadratic forms, MANOVA random effects model, and robustness studies involving mixtures of multivariate Gaussian distributions~\cite{GN1999}. The distribution of sum of Wishart matrices serves as a natural candidate distribution for modeling realized covariance and is of fundamental importance to the multivariate Behrens-Fisher problem~\cite{TG1983,NM1986,Sheppard2008}. Moreover, it has applications in quantitative finance~\cite{Sheppard2008}, telecommunication~\cite{NAH2011,CNSS2003}, sensor network related algorithms~\cite{REM2011}, etc.

The sum of independent Wishart matrices taken from distributions with identical covariance matrices gives rise, again, to a Wishart distribution with the same covariance matrix~\cite{Anderson2003,AHSE1995}; see Eq.~\eqref{CorrWish} ahead. However, for the case of unequal covariance matrices, deriving the distribution of the sum of Wishart matrices becomes extremely difficult and impractical. Even in the case of two Wishart matrices, the distribution of sum involves a hypergeometric function with matrix arguments~\cite{GN1999}. This complicated and rotationally-noninvariant nature of the matrix distribution makes the evaluation of statistics of eigenvalues an intractable task. 

In the present work we take the first steps towards solving this problem and consider the sum of two independent complex Wishart matrices associated with unequal covariance matrices, such that one of the covariance matrices is proportional to the identity matrix, while the second one is arbitrary. To tackle this problem we employ a generalization of the Harish-Chandra-Itzykson-Zuber unitary-group integral~\cite{James1964,Orlov2004}. We derive compact results for the joint probability density of eigenvalues, as well as the marginal density which involves easily evaluable determinantal structure. The analytical predictions are verified by numerical simulations, and we find excellent agreements.

% SECTION 2
\section{Distribution of the sum of two complex Wishart matrices}
Let us consider two independent complex matrices $A$ and $B$ of dimensions $n\times n_A$ and $n\times n_B$ taken, respectively, from the distributions
\begin{equation}
\mathcal{P}_A(A)=(\pi^{-n}\det \Sigma_A^{-1})^{n_A} \,e^{-\tr (A^\dag \Sigma_A^{-1} A)},~~~
\mathcal{P}_B(B)=(\pi^{-n}\det \Sigma_B^{-1})^{n_B} \, e^{-\tr(B^\dag \Sigma_B^{-1} B)}.
\end{equation}
Here `$\tr$' and `$\det$' represent the trace and the determinant, respectively, and `$\dag$' denotes the Hermitian-conjugate.  $\Sigma_A$, $\Sigma_B$ are the covariance matrices. We assume that $n_A,n_B\ge n$. We have $\int d[A] \,\mathcal{P}_A(A)=\int d[B]\, \mathcal{P}_B(B)=1$. Here $d[A]=\prod_{j=1}^n \prod_{k=1}^{n_A} d A_{jk}^{(R)}\, d A_{jk}^{(I)}$, with $(R)$ and $(I)$ representing the real and imaginary parts, respectively. Similar definition is to be understood for $d[B]$. Since the domains of $A$ and $B$ remain invariant under unitary rotation, without loss of generality, we may take $\Sigma_A$ and $\Sigma_B$ as diagonal matrices. We consider $\Sigma_A=\diag(\sigma_{A1},...,\sigma_{An})$ and $\Sigma_B=\diag(\sigma_{B1},...,\sigma_{Bn})$.  The matrices $AA^\dag$ and $BB^\dag$ are then $n$-variate complex-Wishart-distributed, i.e.,  $AA^\dag\sim\mathcal{W}_n^{\mathbb{C}}(n_A,\Sigma_A)$ and $BB^\dag\sim\mathcal{W}^{\mathbb{C}}_n(n_B,\Sigma_B)$; $n_A, n_B$ being the respective degrees of freedom. 

We are interested in the statistics of the ensemble of $n\times n$ dimensional Hermitian matrices
\begin{equation}
H=AA^\dag + BB^\dag.
\end{equation}
The distribution of $H$ can be obtained as
 \begin{equation}
\mathcal{P}_H(H)=\int \!d[A]\int\! d[B] \delta(H-AA^\dag -BB^\dag) \mathcal{P}_A(A) \mathcal{P}_B(B).
\end{equation}
The delta function with matrix argument in the above equation represents the product of delta functions with scalar arguments, one for each independent real and imaginary component of $H-AA^\dag -BB^\dag$.
Using the Fourier representation for delta function we can write
\begin{equation}
\mathcal{P}_H(H)\propto \int d[K]\int d[A]\int d[B] e^{i \tr (K(H-AA^\dag -BB^\dag ))} 
 e^{-\tr (A^\dag \Sigma_A^{-1} A)} \,e^{-\tr(B^\dag \Sigma_B^{-1} B)}.
\end{equation}
Here $K$ is an $n\times n$ dimensional matrix with the same symmetry properties as $H-AA^\dag -BB^\dag$, i.e., it is Hermitian. The Gaussian integrals over $A$ and $B$ can be performed trivially and result in 
\begin{equation}
\label{PHK}
\mathcal{P}_H(H)\propto \int d[K] e^{i \tr (K H)}\det\!\! \,^{-n_A}(\Sigma_A^{-1}+i K)
\det\!\!\,^{-n_B}(\Sigma_B^{-1}+i K).
\end{equation}
As shown in the appendix, this can be brought to the form
\begin{equation}
\label{PHF}
\mathcal{P}_H(H)\propto \det\!\!\,^m H\,\, e^{-\tr(\Sigma_A^{-1} H)} F(H),
\end{equation}
where $m=n_A+n_B-n$, and $F(H)$ is the following matrix integral involving the Jacobi ensemble:
\begin{equation}
F(H)=\int_0^{\1} \!\!d[T]\det\!\,^{n_A-n}(\1-T)\,\det\!\,^{n_B-n}T 
 e^{\tr ((\Sigma_A^{-1}-\Sigma_B^{-1})HT)}.
\end{equation}
Here $T$ is an $n\times n$ dimensional Hermitian matrix. If the covariance matrices happen to be equal, i.e., $\Sigma_A=\Sigma_B=\Sigma$, then $F(H)$ gives just a constant and we obtain 
\begin{equation}
\label{CorrWish}
\mathcal{P}_H(H)\propto \det\!\!\,^{m} H\, \,e^{-\tr (\Sigma^{-1} H)},
\end{equation}
showing that $H$ is complex-Wishart-distributed as $\mathcal{W}^{\mathbb{C}}_n(n_A+n_B,\Sigma)$~\cite{Anderson2003,AHSE1995}. Exact as well as asymptotic results for various eigenvalue statistics are known for this case~\cite{SMM2006,ZCW2009,VP2010,RKG2010,DM2011,RKGZ2012,WG2013,Forrester2013,WG2014}.

In the general case $F(H)$ can be represented in terms of a confluent Hypergeometric function of matrix argument~\cite{Macdonald1987, EK2014},
\begin{eqnarray}
\label{FH}
\nonumber
F(H)\propto \,_1F_1(n_B;n_A+n_B;(\Sigma_A^{-1}-\Sigma_B^{-1})H)~~~~~~~~~~~~\\
=e^{\tr((\Sigma_A^{-1}-\Sigma_B^{-1}) H)}\,_1F_1(n_A;n_A+n_B;(\Sigma_B^{-1}-\Sigma_A^{-1})H).
\end{eqnarray}
The second line in the above equation follows from the Kummer's transformation~\cite{James1964,Macdonald1987}. Therefore, we obtain the distribution of $H$ as
\begin{eqnarray}
\label{PH}
\begin{matrix}
\mathcal{P}_H(H)&=&\mathcal{C} \,\det\!\!\,^m H\, e^{-\tr(\Sigma_A^{-1} H)} \,_1F_1(n_B;n_A+n_B;(\Sigma_A^{-1}-\Sigma_B^{-1})H)~~\\
&=&\mathcal{C} \,\det\!\!\,^m H\, e^{-\tr(\Sigma_B^{-1} H)}  \,_1F_1(n_A;n_A+n_B;(\Sigma_B^{-1}-\Sigma_A^{-1})H).~~~
\end{matrix}
\end{eqnarray}
 The normalization can be fixed by keeping track of all the constants from the beginning~\cite{SKU}. We have
\begin{equation}
\mathcal{C}^{-1}=\frac{\pi^{n(n-1)/2}}{(\det \Sigma_A^{-1})^{n_A}(\det \Sigma_B^{-1})^{n_B}}\prod_{j=1}^n\Gamma(m+j),
\end{equation}
where $\Gamma(a)$ is the Gamma function. Eq.~\eqref{PH} constitutes one of the key results of this paper. In the case of identical covariance matrices $\,_1F_1(n_B;n_A+n_B;(\Sigma_A^{-1}-\Sigma_B^{-1})H)$ gives 1, and thereby we recover Eq.~\eqref{CorrWish}. We remark that the distribution of $H$ in the case of real matrices can also be obtained using the same procedure.

 %SECTION 3
\section{Statistics of eigenvalues}
We now specialize to the case when one of the covariance matrices is proportional to the identity matrix, say $\Sigma_A=\sigma_A \1$, while the second, $\Sigma_B$, is arbitrary. Equivalently, we may consider $\Sigma_B=\sigma_B\1$ and an arbitrary $\Sigma_A$ in the second expression in Eq.~\eqref{PH}. For the former choice, the factor before the Hypergeometric function in Eq.~\eqref{PH} becomes unitarily invariant~\cite{Mehta2004}. Using the eigenvalue-decomposition $H=U^\dag \Lambda U$, where $\Lambda$ is the diagonal matrix with the eigenvalues of $H$, we obtain 
\begin{eqnarray}
\label{PL2}
\nonumber
P(\lambda_1,...,\lambda_n)\propto \Delta_n^2(\{\lambda \}) \prod_{l=1}^n \lambda_l^m\, e^{-\sigma_A^{-1} \lambda_l} \int_{\mathbb{U}_n} d\mu(U) \\
\times \,_1F_1(n_B;n_A+n_B;(\sigma_A^{-1}\1-\Sigma_B^{-1})\,U^\dag\Lambda U).
\end{eqnarray}
Here $\Delta_n(\{\lambda\})=\prod_{j>k}(\lambda_j-\lambda_k)$ is the Vandermonde determinant and $d\mu(U)$ represents the Haar measure over the unitary group $\mathbb{U}_n$. The above group integral can be performed using the result below, and leads to a Hypergeometric function of two matrix arguments~\cite{James1964,Orlov2004},
\begin{equation}
\label{GInt}
 \int_{\mathbb{U}_n} d\mu(U) \,_1F_1(a;b;X U^\dag Y U)
 =\,_1\mathcal{F}_1(a;b;X,Y).~~~
\end{equation}
This result is a generalization of the celebrated Harish-Chandra-Itzykson-Zuber unitary group integral.
We have the following representation for $\,_1\mathcal{F}_1(a;b;X,Y)$ in terms of a determinant involving the eigenvalues $\{x_1,...,x_n\}$ and $\{y_1,...,y_n\}$ of normal matrices $X$ and $Y$~\cite{Orlov2004}:
 \begin{equation}
 \label{DetRep}
_1\mathcal{F}_1(a;b;X,Y)\propto\frac{\det\left[_1F_1(a-n+1;b-n+1;x_j y_k)\right]}{\Delta_n(\{x\})\Delta_n(\{y\})},\\
\end{equation}
where  $_1F_1$ inside the determinant is the usual confluent hypergeometric function with scalar arguments.
Using Eqs.~\eqref{GInt} and~\eqref{DetRep} in Eq.~\eqref{PL2}, we obtain the joint probability density of the eigenvalues of $H$ as
\begin{equation}
\label{JPD}
P(\lambda_1,...,\lambda_n)=C\, \Delta_n (\{\lambda \}) \prod_{l=1}^n  \lambda_l^m e^{-\sigma_A^{-1} \lambda_l} 
\det\big[\,_1F_1(\alpha;\, \gamma;\, (\sigma_A^{-1}-\sigma_{Bj}^{-1}) \lambda_k \big]_{j,k=1,...,n}.
\end{equation}
Here $C$ is the normalization constant, and $\alpha=n_B-n+1, \gamma=n_A+n_B-n+1.$

It is worth mentioning that the confluent hypergeometric function in Eq.~\eqref{JPD} can be represented in terms of more elementary functions. Noting that $\gamma=\alpha+n_A$, we have
$$
\,_1 F_1(\alpha;\, \gamma;\,z)=\frac{(-z)^{-\alpha}}{\mathds{B}(\alpha,n_A)}\sum_{k=0}^{n_A}z^{-k}\binom{n_A-1}{k}\,\text{\ipagamma}(\alpha+k,-z),\\
$$
where $\binom{a}{b}$ represents the binomial coefficient, and $\mathds{B}(a,b)$ and \ipagamma$(a,b)$ are the Beta function and the lower incomplete gamma function, respectively. 
This simplifies further for special cases or parameter values. For instance, $n_B=n$ gives
\begin{eqnarray*}
\,_1 F_1(1;n_A+1\,;\,z)=n_A\, z^{-n_A}\, e^z\,\text{\ipagamma}(n_A,z)\\
=n_A!\,\,z^{-n_A}\Big(e^z-\sum_{k=0}^{n_A-1}\frac{z^k}{k!}\Big),
\end{eqnarray*}
which also includes the case $n_A=n_B=n$.

To evaluate the normalization constant $C$ in Eq.~\eqref{JPD}, we expand the Vandermonde determinant as well as the determinant involving the hypergeometric functions and perform the integral over the eigenvalues using the relation
\begin{equation}
\label{Int}
\int_0^\infty\!\!d\lambda\,  \lambda^\mu e^{-s \lambda} \,_1F_1(a; b; c \lambda)=\frac{\Gamma(\mu+1)}{s^{\mu+1}}\!\,_2F_1\Big(a;\mu+1;b;\frac{c}{s}\Big),
\end{equation}
which holds whenever the integral is convergent. The expression obtained afterwards can be reformulated as a determinant~\cite{SKU}. We obtain
\begin{equation}
\label{Cinv}
C^{-1}=n! \,\sigma_A^{nm+n(n+1)/2}\prod_{l=1}^n\Gamma(m+l)\det\left[\,_2F_1(\alpha,m+j,\gamma,1-\sigma_A\sigma_{Bk}^{-1})\right]_{j,k=1,...,n},
\end{equation}
such that $\int_0^\infty d\lambda_1\cdots \int_0^\infty d\lambda_n P(\lambda_1,...,\lambda_n)=1.$ 

When the $\sigma_{Bk}$'s have multiplicity greater than 1, i.e., if some or all of the $\sigma_{Bk}$'s are identical, then the determinants in Eqs.~\eqref{JPD} and \eqref{Cinv} become zero. In such \emph{degenerate} cases the appropriate result can be obtained by a limiting procedure. Eq.~\eqref{JPD} is another important contribution of this work. Fig. 1 shows the joint probability density $P(\lambda_1,\lambda_2)$ corresponding to the $n=2$ case, with parameter values as indicated in the caption. The agreement between the analytical result and numerical-simulation result is excellent.
% FIGURE 1
\begin{figure*}[!t]
\centering
\begin{subfigure}{.54\textwidth}
  \centering
  \includegraphics[width=\linewidth]{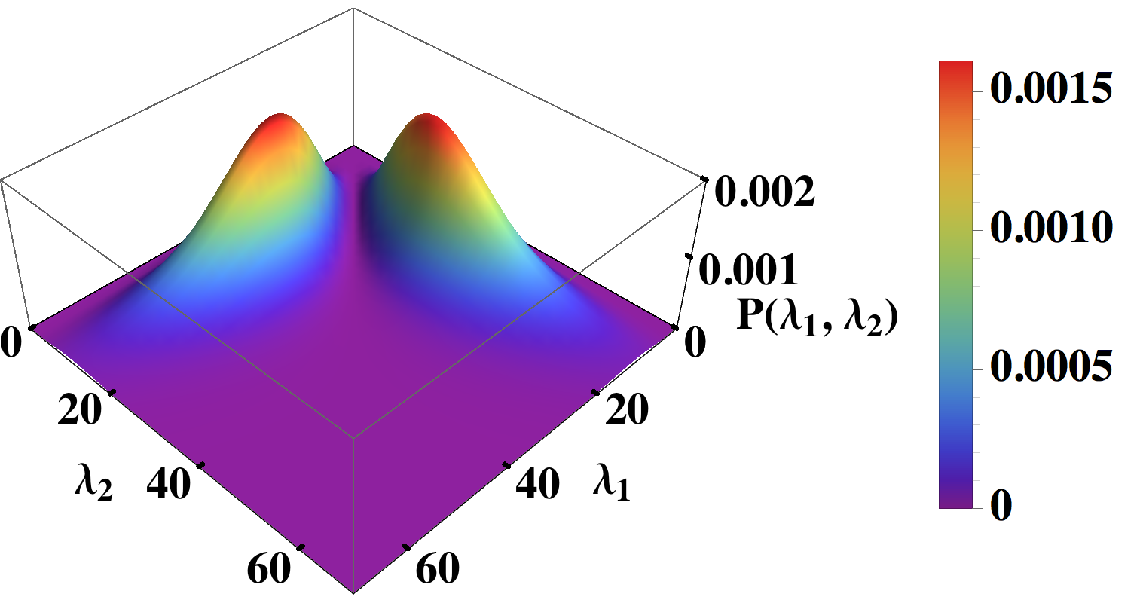}
  \caption{}
  \label{fig:sub1}
\end{subfigure}
\begin{subfigure}{.41\textwidth}
  \centering
  \includegraphics[width=\linewidth]{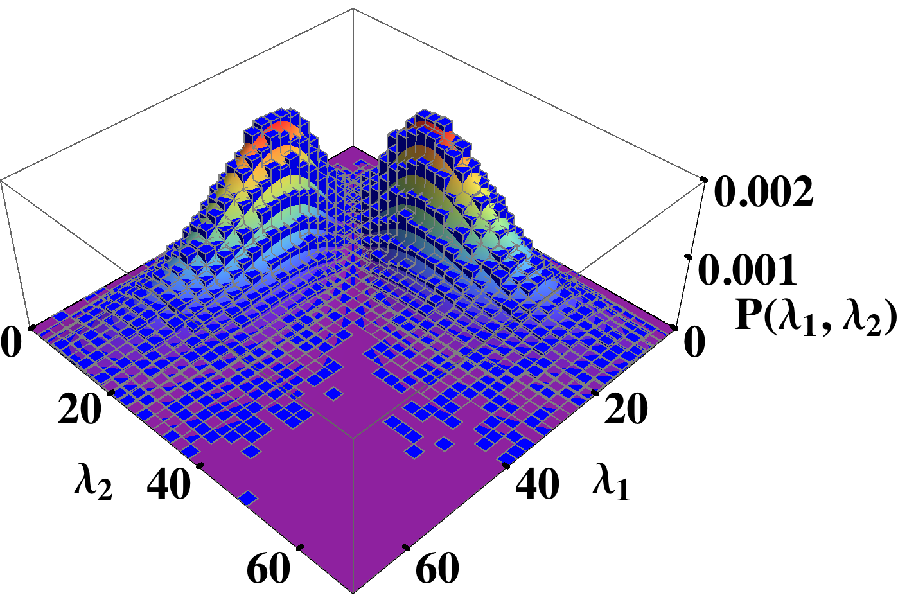}
  \caption{}
  \label{fig:sub2}
\end{subfigure}
\caption{(a) Analytical joint probability density of eigenvalues for $n=2, n_A=3,n_B=4$, and $(\sigma_A; \sigma_{B1}, \sigma_{B2})=(1; 4, 6)$, (b) Analytical plot overlaid on the histogram obtained ~~~~\\ from numerical simulation.~~~~~~\hspace{9.8cm}~~~~}
\label{Fig1}
\end{figure*}
% FIGURE 1%

 We remark that the joint probability density given by Eq.~\eqref{JPD} is of the form of a bi-orthogonal ensemble in the sense of Borodin~\cite{Borodin1998}. Such a structure, in view of the results in~\cite{Borodin1998}, implies existence of compact expression for the $n$-point correlation function~\cite{Mehta2004},
 \begin{equation}
R_n(\lambda_1,...,\lambda_n)=\frac{N!}{(N-n)!}\int_0^\infty d\lambda_{n+1}\cdots\int_0^\infty d\lambda_{N}P(\lambda_1,...,\lambda_n),
 \end{equation} 
 which includes the level density $R_1(\lambda)$.
% FIGURE 2
\begin{figure}[!h]
  \centering
    \includegraphics[width=0.7\textwidth]{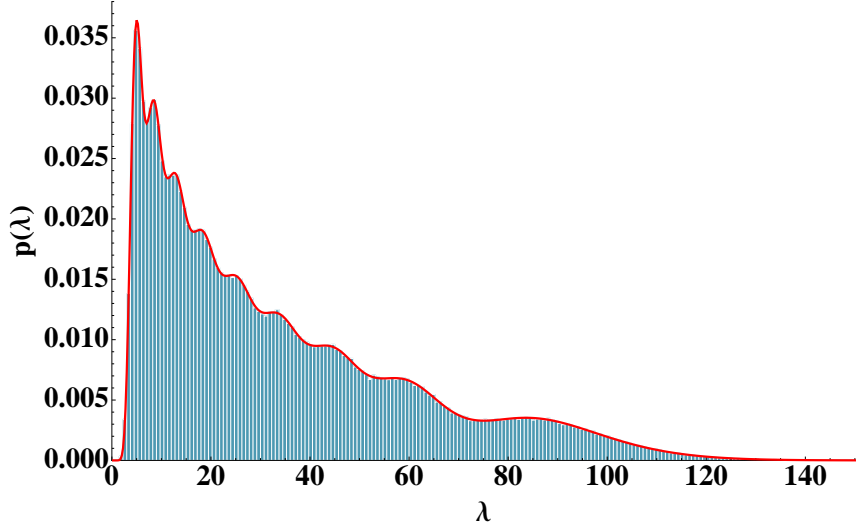}
    \caption{Marginal \,density\, of \,eigenvalues \,for $n=9,n_A=10,n_B=13$,~ and~~~~~~~~~\\ $(\sigma_{A}; \sigma_{B1},...,\sigma_{B9})=(2; 4/17, 1/2, 5/3, 2/7, 6/5, 10/29, 5/16, 9/11, 4)$. The histogram\\ is from the numerical simulation while the solid line is the analytical prediction.~~}
    \label{Fig2}
\end{figure}
% FIGURE 2
We now move on to calculate the marginal density $p(\lambda)$ of eigenvalues, which is given by
\begin{equation}
\label{MDens}
p(\lambda)=\int_0^\infty d\lambda_2\cdots\int_0^\infty d\lambda_n\,P(\lambda, \lambda_2...,\lambda_n),
\end{equation}
and is related to the level density as $p(\lambda)=R_1(\lambda)/N$.
To this end, we expand the determinants in Eq.~\eqref{JPD} and then integrate over the $n-1$ eigenvalues with the aid of Eq.~\eqref{Int}. The resulting expression can be recast in terms of the determinant of an $n+1$-dimensional matrix~\cite{RKGZ2012,SKU}. We have
\begin{equation}
\label{pL}
p(\lambda)=c \,\lambda^m e^{-\sigma_A^{-1} \lambda}
\det\begin{bmatrix} 
0 & [f_k(\lambda)]_{k=1,...,n} \\
[g_j(\lambda)]_{j=1,...,n} & [h_{j,k}]_{j,k=1,...,n}
\end{bmatrix};
\end{equation}
\begin{eqnarray*}
&& f_k(\lambda)=\,_1F_1(\alpha;\, \gamma;\, (\sigma_A^{-1}-\sigma_{Bk}^{-1}) \lambda),\\ 
&& g_j(\lambda)=\lambda^{j-1}/\Gamma(m+j),\\
&& h_{j,k}=\sigma_A^{m+j}\,_2F_1(\alpha;\, m+j;\, \gamma; 1-\sigma_A \sigma_{Bk}^{-1}).
\end{eqnarray*}
To enunciate the notation used above we consider, as an example, the $n=2$ case and write the determinant part explicitly:
$$
\det\begin{bmatrix} 
0 & f_1(\lambda) & f_2(\lambda) \\
g_1(\lambda) & h_{1,1} & h_{1,2} \\
g_2(\lambda) & h_{2,1} & h_{2,2}
\end{bmatrix}.
$$

The normalization $c$ in Eq.~\eqref{pL} is given by
\begin{equation}
c^{-1}=-n \det[h_{j,k}]_{j,k=1,...,n}.
\end{equation}
Eq.~\eqref{pL} constitutes the main result of this paper. Fig.~\ref{Fig2} shows an example where we compare the analytical and simulation results. The parameter values are indicated in the caption. We find perfect agreement.

% FIGURE 3
\begin{figure}[!t]
  \centering
    \includegraphics[width=0.7\textwidth]{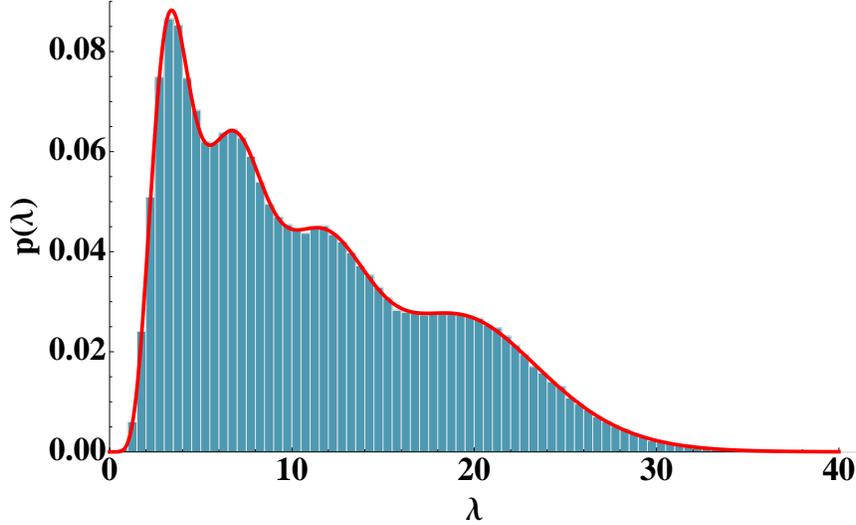}
    \caption{Marginal density of eigenvalues for a {\it degenerate} case: $n=4,n_A=6,n_B=5$, and $(\sigma_{A}; \sigma_{B1},...,\sigma_{B4})=(5/4; 5/7, 5/7, 5/7, 5/7).\hspace{8.2cm}$}
\label{Fig3}
\end{figure}
% FIGURE 3
Again, if some or all of the $\sigma_{Bk}$'s are identical, then we have to take the limit properly to obtain the appropriate expression. For instance, if all the $\sigma_{Bk}$'s are equal, viz. $\sigma_{B1}=\cdots =\sigma_{Bn}=\sigma_B$, then $p(\lambda)$ is still given by Eq.\eqref{pL}, but with the following modification~\cite{SKU}:
\begin{eqnarray*}
&& f_k(\lambda)=\lambda^{k-1}\,_1F_1^{(k-1)}(\alpha;\, \gamma;\, (\sigma_A^{-1}-\sigma_B^{-1}) \lambda),\\ 
&& g_j=\lambda^{j-1}/\Gamma(m+j),\\
&& h_{j,k}=\sigma_A^{m+j+k-1}\,_2F_1^{(k-1)}(\alpha;\, m+j;\, \gamma; 1-\sigma_A\sigma_B^{-1}).
\end{eqnarray*}
Here $\,_1F_1^{(k-1)}(\alpha;\, \gamma;\, z)$= $\partial^{k-1} \,_1F_1(\alpha;\, \gamma;\, z)/\partial z^{k-1}$. Similar definition is to be understood for $\,_2F_1^{(k-1)}(\alpha;\, m+j;\, \gamma; z)$. We note that the following relations hold for the $k$th derivative of the confluent and Gauss hypergeometric functions with respect to the last argument:
\vspace{0.3cm}
\begin{eqnarray*}
&&\frac{\partial ^k \,_1F_1(a;b;z)}{\partial z^k}=\frac{(a)_k}{(b)_k}\,_1F_1(a+k;b+k;z),\\
&&\frac{\partial ^k \,_2F_1(a;b;c;z)}{\partial z^k}=\frac{(a)_k (b)_k}{(c)_n}\,_2F_1(a+k;b+k;c+k;z),
\end{eqnarray*}
\noindent
where $(a)_k$ etc. represent the Pochhammer symbol with definition $(a)_k=\Gamma(a+k)/\Gamma(k)$. 
In Fig.~\ref{Fig3} we consider a \emph{degenerate} case where both the covariance matrices are proportional to the identity matrix. Once again the analytic and the simulation results agree perfectly.

%SECTION 4
\section{Summary and Discussion}
We considered the problem of computing the eigenvalue statistics of sum of two independent complex Wishart matrices taken from distributions with unequal covariance matrices. We found a complete solution to the problem when one of the covariance matrices is proportional to the identity matrix. We derived a compact result for the joint probability density of eigenvalues which can be used to evaluate the statistics of any observable dependent on the eigenvalues. We also derived an easily computable determinantal expression for the marginal density of eigenvalues. These expressions can be readily implemented in {\sc Mathematica}~\cite{Mathematica}. Finally, we performed numerical simulations to test the analytical results and found perfect agreement.

It remains to see if some compact form can be obtained for the case when both the covariance matrices are arbitrary. Moreover, it will be of interest to explore if the problem involving the sum of more than two Wishart matrices is analytically surmountable, and if there is some underlying deeper structure.

%SECTION 5
\section*{Appendix}
We outline here the steps leading to Eq.~\eqref{PHF}, starting from Eq.~\eqref{PHK}. We introduce another delta-function in Eq.~\eqref{PHK}, involving a new Hermitian matrix $\widetilde{K}$ and afterwards separate the `$K$' and `$\widetilde{K}$' parts:
\begin{eqnarray*}
&&\mathcal{P}_H(H)\propto \int d[K]\int  d[\widetilde{K}] \delta(\widetilde{K}-K) e^{i\tr (K H)}\det\!\!\,^{-n_A}(\Sigma_A^{-1}+i K)\det\!\!\,^{-n_B}(\Sigma_B^{-1}+i K)\\
&&=\int d[K]\int d[\widetilde{K}] \delta(\widetilde{K}-K)e^{i\tr (K H/2)}\det\!\!\,^{-n_A}(\Sigma_A^{-1}+i K)  e^{i \tr (\widetilde{K} H/2)}\det\!\!\,^{-n_B}(\Sigma_B^{-1}+i \widetilde{K})\\
&&= \int d[S]\int d[K]\int d[\widetilde{K}] e^{i\tr (S(\widetilde{K}-K))} e^{i \tr (K H/2)}\det\!\!\,^{-n_A}(\Sigma_A^{-1}+i K)e^{i\tr (\widetilde{K} H/2)}\det\!\!\,^{-n_B}(\Sigma_B^{-1}+i \widetilde{K}).
\end{eqnarray*}
In the last step above we introduced the Fourier representation for $\delta(\widetilde{K}-K)$ with the aid of a Hermitian matrix $S$.
We now consider the transformations $K\rightarrow \Sigma_A^{-1} K$ and $\widetilde{K}\rightarrow\Sigma_B^{-1}\widetilde{K}$. The resulting Jacobians can be absorbed in the overall constant and therefore we obtain
\begin{eqnarray*}
\mathcal{P}_H(H)\!\propto\! \int \!d[S]\!\int \!d[K]  e^{i \tr (K (H/2-S)\Sigma_A^{-1})}\det\,\!\!^{-n_A}(\1+i K)\\
 \times\int \!d[\widetilde{K}]e^{i \tr(\widetilde{K} (H/2+S)\Sigma_B^{-1})}\det\!\!\,^{-n_B}(\1+i \widetilde{K}).
\end{eqnarray*}
The $K$ and $\widetilde{K}$ integrals can be performed using the Ingham-Siegel type integral~\cite{Fyodorov2002}, yielding
\begin{eqnarray*}
\mathcal{P}_H(H)\propto\int d[S] \det\!\!\,^{n_A-n}(H/2-S)\,\,e^{-\tr((H/2-S)\Sigma_A^{-1})}\det\!\!\,^{n_B-n}(H/2+S)\,\,e^{-\tr((H/2+S)\Sigma_B^{-1})}\\
\times\Theta((H/2-S)\Sigma_A^{-1})~\Theta((H/2+S)\Sigma_B^{-1}).
\end{eqnarray*}
Here $\Theta(G)$ represents the matrix theta function, and requires the matrix $G$ to be positive definite ($G>0$) for a non-vanishing result.
Employing the transformation $S\rightarrow (H/2)S$, and observing that $\Sigma_A^{-1}>0, \Sigma_B^{-1}>0, H>0$, we obtain~\cite{Serre2010}
\begin{eqnarray*}
\mathcal{P}_H(H)\propto(\det H)^{(n_A+n_B-n)} e^{-\tr((\Sigma_A^{-1}+\Sigma_B^{-1})H/2)}
\int d[S]\det\,^{n_A-n}(\1-S)\det\,^{n_B-n}(\1+S)\\
\times e^{\tr((\Sigma_A^{-1}-\Sigma_B^{-1})HS/2)}\Theta(\1-S)\Theta(\1+S).
\end{eqnarray*}
The matrix theta functions in the above expression restricts the domain of $S$ in the integration to $-\1<S<\1$.
Finally, introducing the Hermitian matrix $T=(\1+S)/2$ we have
\begin{eqnarray*}
\mathcal{P}_H(H)\propto(\det H)^{(n_A+n_B-n)} e^{-\tr (\Sigma_A^{-1} H)}\int_0^{\1}\!\!\!d[T] \det\!\!\,^{n_A-n}(\1-T)\det\!\!\,^{n_B-n}T \,e^{\tr((\Sigma_A^{-1}-\Sigma_B^{-1})HT)},
\end{eqnarray*}
and hence Eq.~\eqref{PHF}.

% REFERENCES


\begin{thebibliography}{99}

\bibitem{Wishart1928} J. Wishart, \href{http://dx.doi.org/10.1093/biomet/20A.1-2.32}{Biometrika {\bf 20A}, 32 (1928)}.
% WISHART

\bibitem{James1964} A. T. James, \href{http://dx.doi.org/10.1214/aoms/1177703550}{Ann. Math. Statist. {\bf 35}, 475 (1964)}.
% Unitary group integral

\bibitem{Muirhead2009} R. J. Muirhead, {\it Aspects of multivariate statistical theory} Vol. 197 (John Wiley \& Sons, 2009).
% Multivariate statistics 

\bibitem{Anderson2003} T. W. Anderson, {\it An Introduction to Multivariate Statistical Analysis} (John Wiley \& Sons, 2003), 3rd ed.
% Multivariate Statistics and Sum of Wisharts

\bibitem{AHSE1995} H. H. Andersen, M. H{\o}jbjerre, D. S{\o}rensen and P. S. Eriksen, {\it Linear and Graphical Models for the Multivariate Complex Normal Distribution}, Lecture Notes in Statistics, Vol. 101 (Springer-Verlag, New York 1995).
% Multivariate Statistics and Sum of Wisharts

\bibitem{GN1999}  A. K. Gupta and D. K. Nagar, {\it Matrix variate distributions}, Vol. 104 (CRC Press, 1999).
% Multivariate statistics 

\bibitem{LCBM1999} L. Laloux, P. Cizeau, J.-P. Bouchaud, and M. Potters, \href{http://dx.doi.org/10.1103/PhysRevLett.83.1467}{Phys. Rev. Lett. {\bf 83}, 1467 (1999)}.
% Finance

\bibitem{PGRAGS2002} V. Plerou, P. Gopikrishnan, B. Rosenow, L. A. N. Amaral, T. Guhr, and H. E. Stanley, \href{http://dx.doi.org/10.1103/PhysRevE.65.066126}{Phys. Rev. E {\bf 65}, 066126 (2002)}.
% Finance

\bibitem{AFV2010} G. Akemann, J. Fischmann, and P. Vivo, \href{http://dx.doi.org/10.1016/j.physa.2010.02.026}{Physica A {\bf 389}, 2566 (2010)}. 
% Finance

\bibitem{SCSG2014} T. A. Schmitt, D. Chetalova, R. Sch\"{a}fer, and T. Guhr, \href{http://dx.doi.org/10.1209/0295-5075/105/38004}{Europhys. Lett. {\bf 105}, 38004 (2014)}.
% Finance

\bibitem{Detal2011} V. Dahirel {\it et al.}, \href{http://dx.doi.org/10.1073/pnas.1105315108}{Proc. Natl. Acad. Sci. U.S.A. {\bf 108}, 11530 (2011)}.
% HIV

\bibitem{Telatar1999} I. E. Telatar, \href{http://dx.doi.org/10.1002/ett.4460100604}{Eur. Trans. Telecommun. {\bf 10}, 585 (1999)}.
% Communication

\bibitem{TV2004} A. M. Tulino and S. Verdu, {\it Random Matrix Theory and Wireless Communications}, Foundations and Trends Com. and Inf. Th. (now Publishers Inc, Boston, Delft, 2004).
% Communication

\bibitem{SMM2006}  S. H. Simon, A. L. Moustakas, and L. Marinelli, \href{http://dx.doi.org/10.1109/TIT.2006.885519}{IEEE Trans. Inf. Theory {\bf 52}, 5336 (2006)}.
% Communication & Correlated Wishart eigenvalue statistics

\bibitem {ZCW2009} A. Zanella, M. Chiani, and M. Z. Win, \href{http://dx.doi.org/10.1109/TCOMM.2009.04.070143}{IEEE Trans. Commun. {\bf 57} 1050 (2009)}.
% Communication & Correlated Wishart eigenvalue statistics

\bibitem{KP2010a} S. Kumar and A. Pandey, \href{http://dx.doi.org/10.1109/TIT.2010.2044060}{IEEE Trans. Inf. Theory {\bf 56}, 2360 (2010)}.
% Communication

\bibitem{KP2011a} S. Kumar and A. Pandey,  \href{http://dx.doi.org/10.1016/j.aop.2011.04.013}{Ann. Phys. (N.Y.) {\bf 326}, 1877 (2011)}.
% Communication etc.

\bibitem{SV1993} E.V. Shuryak and J. J. M. Verbaarschot, \href{http://dx.doi.org/10.1016/0375-9474(93)90098-I}{Nucl. Phys. A {\bf 560}, 306 (1993)}.
% QCD Dirac operator

\bibitem{Verbaarschot1994} J. Verbaarschot, \href{http://dx.doi.org/10.1103/PhysRevLett.72.2531}{Phys. Rev. Lett. {\bf 72}, 2531 (1994)}.
% QCD Dirac operator

\bibitem{GW1997} T. Guhr and T. Wettig, \href{http://dx.doi.org/10.1016/S0550-3213(97)00556-7}{Nucl. Phys. B {\bf 506}, 589 (1997)}. 
% QCD Dirac operator

\bibitem{VW2000} J. J. M. Verbaarschot and T. Wettig, \href{http://10.1146/annurev.nucl.50.1.343}{Ann. Rev. Nucl. Part. Sci. {\bf 50}, 343 (2000)}.
% QCD Dirac operator review

\bibitem{Akemann2011} G. Akemann, \href{http://dx.doi.org/10.5506/APhysPolB.42.901}{Acta Phys. Pol. B {\bf 42}, 0901 (2011)}. 
% QCD Dirac operator

\bibitem{ZS2001} K. Zyczkowski and H.-J. Sommers, \href{http://dx.doi.org/10.1088/0305-4470/34/35/335}{J. Phys. A {\bf 34}, 7111 (2001)}.
% Entanglement

\bibitem{BL2002} J. N. Bandyopadhyay and A. Lakshminarayan, \href{http://dx.doi.org/10.1103/PhysRevLett.89.060402}{Phys. Rev. Lett. {\bf 89}, 060402 (2002)}.
% Entanglement

\bibitem{NM2010} C. Nadal, S. N. Majumdar, and M. Vergassola, \href{http://dx.doi.org/10.1103/PhysRevLett.104.110501}{Phys. Rev. Lett. {\bf 104}, 110501 (2010)}.
% Entanglement

\bibitem{KP2011b} S. Kumar and A. Pandey, \href{http://dx.doi.org/10.1088/1751-8113/44/44/445301}{J. Phys A. {\bf 44}, 445301 (2011)}.
% Entanglement

\bibitem{VZ2012} Vinayak and M. \v{Z}nidari\v{c}, \href{http://dx.doi.org/10.1088/1751-8113/45/12/125204}{J. Phys. A {\bf 45}, 125204 (2012)}.
% Entanglement

\bibitem{FH1994} P. J. Forrester and T. D. Hughes, \href{http://dx.doi.org/10.1063/1.530639}{J. Math. Phys. {\bf 35}, 6736 (1994)}.
% Mesoscopic

\bibitem{SN1994} K. Slevin and T. Nagao, \href{http://dx.doi.org/10.1103/PhysRevB.50.2380}{Phys. Rev. B {\bf 50}, 2380 (1994)}.
% Mesoscopic

\bibitem{HMMCBF2000} N. S. Holter, M. Mitra, A. Maritan, M. Cieplak, J. R. Banavar, and N. V. Fedoroff, \href{http://dx.doi.org/10.1073/pnas.150242097}{Proc. Natl. Acad. Sci. U.S.A. {\bf 97}, 8409 (2000)}.
% Gene expression data

\bibitem{ABB2000} O. Alter, P. O. Brown, and D. Botstein, \href{http://dx.doi.org/10.1073/pnas.97.18.10101}{Proc. Natl. Acad. Sci. U.S.A. {\bf 97} 10101 (2000)}.
% Gene expression data

\bibitem{DE2002} I. Dumitriu and A. Edelman, \href{http://dx.doi.org/10.1063/1.1507823}{J. Math. Phys. 43, 5830 (2002).}
% Beta ensembles

\bibitem{Forrester2006} P. J. Forrester, \href{http://dx.doi.org/10.1088/0305-4470/39/22/004}{J. Phys. A {\bf 39}, 6861 (2006)}.
% Jacobi ensemble

\bibitem{KP2010b} S. Kumar and A. Pandey, \href{http://dx.doi.org/10.1088/1751-8113/43/8/085001}{J. Phys. A {\bf 43}, 085001 (2010)}.
% Jacobi ensemble

\bibitem{DFS2013} R. Dar, M. Feder, and M. Shtaif, \href{http://dx.doi.org/10.1109/TIT.2012.2233860}{IEEE Trans. Inf. Theory {\bf 59}, 2426 (2013).}
% Jacobi ensemble and fiber communication

\bibitem{AIK2013} G. Akemann, J. R. Ipsen, and M. Kieburg, \href{http://dx.doi.org/10.1103/PhysRevE.88.052118}{Phys. Rev. E {\bf 88}, 052118 (2013)}.
% Product of Wisharts

\bibitem{AKW2013} G. Akemann, M. Kieburg, and L. Wei, \href{http://dx.doi.org/10.1088/1751-8113/46/27/275205
}{J. Phys. A {\bf 46}, 275205 (2013)}.
% Product of Wisharts

\bibitem{Khatri1966} C. G. Khatri, \href{http://dx.doi.org/10.1214/aoms/1177699530}{Ann. Math. Statist. {\bf 37}, 468 (1966)}. 
% Sum of Wisharts

\bibitem{TG1983} W. Y. Tan and R. P. Gupta, \href{http://dx.doi.org/10.1080/03610928308828625}{Commun. Statist. - Theory Meth. {\bf 12}, 2589 (1983)}.
% Sum of Wisharts

\bibitem{NM1986} D. G. Nel and C. A. Van Der Merwe, \href{http://dx.doi.org/10.1080/03610928608829342}{Commun. Statist. - Theory Meth. {\bf 15}, 3719 (1986)}.
% Sum of Wisharts

\bibitem{Sheppard2008} K. Sheppard, (unpublished), \url{http://www.kevinsheppard.com/images/e/e2/PSDMEM_Sheppard.pdf}
% Sum of Wisharts

\bibitem{NAH2011} B. Nosrat-Makouei, J. G. Andrews, and R. W. Heath,  \href{http://dx.doi.org/10.1109/TSP.2011.2124458}{IEEE Trans. Sig. Process. {\bf 59}, 2783 (2011)}.
% Sum of Wisharts-Communication

\bibitem{CNSS2003} K. Conradsen, A. A. Nielsen, J. Schou, and H. Skriver, \href{http://dx.doi.org/10.1109/TGRS.2002.808066}{IEEE Trans. Geosci. Remote Sensing {\bf 41}, 4 (2003)}.
% Sum of Wisharts-Communication

\bibitem{REM2011} N. Ramakrishnan, E. Ertin, and R. L. Moses,  \href{http://dx.doi.org/10.1109/JSTSP.2011.2119291}{IEEE J. Sel. Top. Sig. Proces. {\bf 5}, 665 (2011)}.
% Sensor networks

\bibitem{Orlov2004} A. Y. Orlov, \href{http://dx.doi.org/10.1142/S0217751X04020476}{Int. J. Mod. Phys. A {\bf 19}, 276 (2004)}.
% Unitary group integral

\bibitem{VP2010} Vinayak and A. Pandey, \href{http://dx.doi.org/10.1103/PhysRevE.81.036202}{Phys. Rev. E {\bf 81}, 036202 (2010)}.
% Correlated Wishart eigenvalue statistics

\bibitem{RKG2010} C. Recher, M. Kieburg, and T. Guhr, \href{http://dx.doi.org/10.1103/PhysRevLett.105.244101}{Phys. Rev. Lett. {\bf 105}, 244101 (2010)}.
% Correlated Wishart eigenvalue statistics

\bibitem{DM2011} P. Dharmawansa and M. R. McKay, \href{http://dx.doi.org/10.1016/j.jmva.2011.01.004}{J. Multivar. Anal. {\bf 102}, 847 (2011)}.
% Correlated Wishart eigenvalue statistics

\bibitem{RKGZ2012}  C. Recher, M. Kieburg, T. Guhr, and M. R. Zirnbauer, \href{http://dx.doi.org/10.1007/s10955-012-0567-x}{J. Stat. Phys. {\bf 148}, 981 (2012)}. 
% Correlated Wishart eigenvalue statistics

\bibitem{WG2013} T. Wirtz and T. Guhr, \href{http://dx.doi.org/10.1103/PhysRevLett.111.094101}{Phys. Rev. Lett. {\bf 111}, 094101 (2013)}.
% Correlated Wishart eigenvalue statistics

\bibitem{Forrester2013} P. J. Forrester, \href{http://dx.doi.org/10.1142/S2010326313500111}{Random Matrices: Theory Appl. {\bf 02}, 1350011 (2013)}.
% Spiked Wishart eigenvalue statistics

\bibitem{WG2014} T. Wirtz and T. Guhr, \href{http://dx.doi.org/10.1088/1751-8113/47/7/075004}{J. Phys. A {\bf 47}, 075004 (2014)}.
% Correlated Wishart eigenvalue statistics

\bibitem{Macdonald1987} I. G. Macdonald, Hypergeometric functions I (handwritten notes), 1987-1988, \href{http://arxiv.org/abs/1309.4568}{arXiv:1309.4568}
% 1F1 integral representation

\bibitem{EK2014} A. Edelman and P. Koev, \href{http://dx.doi.org/10.1142/S2010326314500099}{Random Matrices: Theory Appl. {\bf 03}, 1450009 (2014)}.
% 1F1 integral representation

\bibitem{SKU} S. Kumar (unpublished).
% Details

\bibitem{Mehta2004} M. L. Mehta, {\it Random Matrices} (Academic Press, New York, 2004), 3rd ed.
% Invariance

\bibitem{Borodin1998} A. Borodin, \href{http://dx.doi.org/10.1016/S0550-3213(98)00642-7}{Nucl. Phys. B {\bf 536}, 704 (1998)}.
% Biorthogonal ensembles

\bibitem{Mathematica} Wolfram Research Inc., {\sc Mathematica} {\it Version 9.0}, Champaign, Illinois (2013).
% Mathematica

\bibitem{Fyodorov2002} Y. V. Fyodorov, \href{http://dx.doi.org/10.1016/S0550-3213(01)00508-9}{Nucl. Phys. B {\bf 621}, 643 (2002)}.
% Ingham-Siegel integral

\bibitem{Serre2010} D. Serre, \it{Matrices: Theory and Applications} (Springer, 2010), 2nd ed.
% Positive definite matrices


\end{thebibliography}
\end{document}